# Exploring the role of sample thickness for hyperspectral microscopy tissue discrimination through Monte Carlo simulations

**Laura Quintana-Quintana**[1*], **Mark Witteveen**[2], **Behdad Dashtbozorg**[2], **Samuel Ortega**[3,1], **Theo J.M. Ruers**[2], **Henricus J.C.M. Sterenborg**[2], **and Gustavo M. Callico**[1]

[1]*Institute for Applied Microelectronics (IUMA), University of Las Palmas de Gran Canaria (ULPGC), Spain*
[2]*Image-Guided Surgery, Department of Surgery, The Netherlands Cancer Institute, Amsterdam, Netherlands*
[3]*Norwegian Institute of Food, Fisheries and Aquaculture Research (Nofima), Tromsø, Norway*
*\*lquintana@iuma.ulpgc.es*

**Abstract:** Recent advancements in multispectral (MS) and hyperspectral (HS) microscopy have focused on sensor and system improvements, yet sample processing remains overlooked. We conducted an analysis of the literature, revealing that 40% of studies do not report sample thickness. Among those that did report it, the vast majority, 98%, used 2–10 μm samples. This study investigates the impact of unstained sample thickness on MS/HS image quality through light transport simulations. Monte Carlo simulations were conducted on various tissue types (i.e., breast, colorectal, liver, and lung). The simulations revealed that thin samples reduce tissue differentiation, while higher thicknesses (approximately 500 μm) improve discrimination, though at the cost of reduced light intensity. These findings highlight the need to study and optimize sample thickness for enhanced tissue characterization and diagnostic accuracy in MS/HS microscopy.

## 1. Introduction

Multispectral (MS) and hyperspectral (HS) imaging (MSI/HSI) are gaining increasing interest in medical applications due to the enhanced information they provide. These techniques capture diffuse spectra that arise from the interaction of light with tissues, which is unique since it depends on the inherent optical properties of the tissue itself. The difference in absorption ($\mu_a$) and scattering ($\mu_s$) coefficients of each tissue attenuates light differently across various wavelengths, providing distinct spectral signatures [1]. Specifically, MSI/HSI in microscopic histological analysis allows spatial and spectral examination of biological specimens in detail [2]. Some applications include the diagnosis of diabetic condition via retinal imaging [3], Alzheimer's disease biomarker identification in plasma samples [4], the detection of head and neck squamous cell carcinoma on histologic slides [5], or the classification of cholangiocarcinoma from pathology images [6].

    MS/HS microscopy is typically performed by attaching a MS or HS camera to a bright-field microscope [7–9], enabling the capture of spectral features from tissues at a fine scale, such as mammalian cells (5 to 25 μm) or red blood cells (6 to 8 μm) [10]. While much of the development of these systems has focused on improving sensors and imaging technology [11–14], sample preparation has mostly followed the traditional methods used in histology analysis. Biopsies are either embedded in paraffin or frozen before being sliced into thin sections using a microtome or cryostat, respectively. These sections are typically cut to a thickness of 2 to 10 μm, matching the scale of individual cells and allowing for the spatial differentiation of structures within the sample [10].

However, in MS/HS microscopy, special attention must be given to sample preparation, as MS/HS technology relies on the interaction of light with the sample, and light behaves differently at small tissue thicknesses compared to bulk tissue. According to diffusion theory, which approximates light propagation in highly scattering media such as tissue, the mean free path, defined as $l_s = 1/\mu_a + 1/\mu_s$, represents the average distance a photon travels through tissue before its direction (scattering) or energy (absorption) is significantly altered [15]. In bulk tissue, the average distance travelled by photons before being absorbed or detected is higher or much higher than the mean free path (diffuse regime). However, in microscopy, if the thickness of a tissue slice is so short that does not exceed one mean free path, the detected photons will have nearly the same direction and energy as the emitted ones (sub-diffuse regime). In tissues, the mean free path typically ranges from 10 to 1000 μm, with around 100 μm being common in the visible spectrum [16]. As a result, when tissue is sliced to a thickness of 2–10 μm for conventional microscopy analysis, if captured without further processing (i.e., unstained), the detected light would be almost identical to the emitted light, resulting in low contrast and making the sample appear almost transparent to the human eye.

Thus, in bright-field microscopy, contrast enhancement of biological samples is performed using dyes (exogenous chromophores), such as hematoxylin and eosin (H&E) staining [17]. Staining techniques are designed to ensure that each structure of the sample absorbs different dyes, such as hematoxylin for the cell nuclei and eosin for the extracellular matrix and cytoplasm. Since each dye absorbs light at distinct wavelengths, contrast is created on the sample [18]. These samples are then examined either visually or by capturing RGB (Red, Green and Blue) images. Nonetheless, staining alters the intrinsic optical properties of tissues, confining spectral contrast to the specific dyes used, which are identical across all samples. As a result, MSI/HSI techniques cannot benefit from the spectral contrast provided by the endogenous chromophores inherent to each tissue type and human being.

Therefore, according to light propagation in tissue, to fully utilize MS/HS microscopy in the broadest context the focus should be on capturing these endogenous chromophores of tissues, which provide richer spectral data not only in the visible spectrum (VIS) but also in the near-infrared range (NIR). While a conventional tissue thickness of around 5 μm may serve this purpose, it may not always be optimal. Increasing thickness provides more detailed spectral signatures, until a certain point where light penetration is reduced, resulting in lower light intensity. To fully leverage the capabilities of MS/HS microscopy it is crucial to pay special attention to sample thickness, balancing the ability to differentiate cells on a sample (2-10 μm) with the challenge of capturing meaningful spectra from the sample itself (~100 μm). Selecting an optimal thickness ensures enough scattering and absorption events to enhance contrast while preserving the visualization of tissue structures.

This study aims to evaluate the impact of sample thickness in MS/HS microscopy spectral data and to establish a benchmark for optimizing tissue thickness in future studies. We begin with a comprehensive review and analysis of the literature to identify the sample thicknesses previously used in MS/HS microscopy studies. Next, a virtual specimen will be designed using a Monte Carlo (MC) Light Transport Simulator [19,20] to further investigate the effects of different sample thicknesses over the diffuse spectra obtained. Using this approach, spectral signatures of normal and lesioned tissues (breast, colorectal, liver, and lung) will be simulated at different sample thicknesses. Finally, spectral contrast will be assessed by evaluating the accuracy of each thickness in distinguishing between normal and lesioned tissue. Thus, this work provides practical insights and guidelines for improving spectral imaging protocol in future studies.

## 2. State of the Art

In the field of medical imaging, the integration of MS/HS microscopy has emerged as a powerful technique for real-time non-ionizing diagnostics. Ortega *et al.*, conducted a systematic review on MS/HS microscopy [2] using the following query: (Hyperspectral OR Multispectral)

AND (histology OR pathology OR histopathology). From 2004-2019, 1776 documents were retrieved, and a rigorous selection process chose 192 documents that met the eligibility criteria established by the authors. Of those, 85 were based on the application of MS/HS microscopy technology for diagnosing various diseases. Building upon this foundation, we undertook an update to the review, extending the search period from August 13th, 2019, to December 12th, 2023. This yielded 759 additional documents, among which 100 met the eligibility criteria established by Ortega *et al.* (humans or mammals samples captured using MSI, HSI, or near-infrared (NIR) sensors coupled to an optical microscope) [2]. These works were meticulously scrutinized to identify and document the specific tissue thickness employed in the different MS/HS microscopy applications.

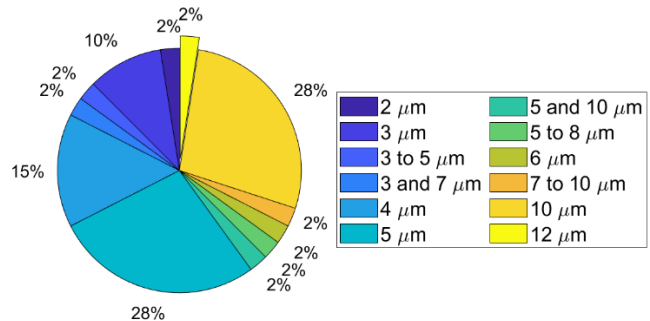

Fig. 1. Tissue thickness distribution in MS and HS microscopy studies from the state-of-the-art, with 5 μm and 10 μm being the most represented thicknesses at 28 % each.

Tissue thickness was not reported as a parameter in 40 % of the eligible studies, highlighting the limited attention given to this crucial aspect in existing literature. For the studies that did include this information (60 documents), the distribution of reported values is shown in Fig. 1. Notably, 5 μm is the most frequently documented tissue thickness, accounting for 28 % of the studies specifying this parameter, since it is the thickness used for routine paraffin sections [21]. Similarly, another 28 % reported a thickness of 10 μm, which corresponds to the maximum cutting capacity of most microtomes. Of these, 4 articles belong to the same research group, investigating normal and diabetic retina of rats [22–25]. They selected 10 μm-thick serial sections that passed through the optic nerve head and dyed them with hematoxylin and eosin (H&E) for HS microscopic examination. Chipala et al. [26] demonstrated that optical density (inversely proportional to transmittance) in H&E stained slides increases with tissue thickness (e.g., rising from 0.2 to 0.5 as thickness ranges from 2 to 10 μm), revealing the impact of thickness on optical properties. A notable exception is the study by Pertzborn *et al*. [27], which utilized 12 μm-thick frozen sections. These are slightly thicker than standard due to cryotome limitations. HSI data from these sections was used to train a 3D convolutional neural network for detecting oral squamous cell carcinoma. In summary, this analysis of the state of the art on MS/HS microscopy highlights the need for further research into how tissue thickness impacts light transmission and spectral contrast for MS/HS microscopy applications, emphasizing its importance for accurate imaging and analysis.

## 3. Materials and Methods

In this section, the MC simulation framework is discussed to examine tissue-light interactions at different thicknesses in both normal and lesioned tissues (specifically breast, lung, colorectal, and liver). To achieve this, a comprehensive analysis of the instrumentation and optics of the HS microscopy system is conducted, along with an investigation of the optical properties of the tissues (absorption, scattering, scattering phase function, and refractive index). Additionally, spectral similarity metrics will be introduced to assess the similarity of the spectral signatures simulated for normal and lesioned tissues at a certain thickness. This approach will help to identify the thickness that most effectively distinguishes normal and lesioned tissue.

*3.1 Microscopic HS System*

The objective of this study is to investigate the influence of sample thickness on spectral signatures. Since there is no standard MS/HS microscope, we will focus on an HS microscope based on a widely used configuration [28,29]. A HS sensor (i.e., the Hyperspec® VNIR -Visual and Near Infrared- A-Series HS camera -HeadWall Photonics, MA, USA-) integrated with a conventional bright-field microscope (i.e., the Olympus BX-53 -Olympus, Tokyo, Japan-) has been used. The system transmits light through a collimator to achieve uniform radiance of the sample, commonly referred to as Köhler illumination. Moreover, the light goes through one of the four objective lenses (10×, Numerical Aperture = 0.3 and Working Distance = 18 mm, HS sensor pixel pitch = 7.4 µm and thus spatial resolution = 0.74 µm) optimized for infrared (IR) observations. Fig. 2 depicts a schematic of the tissue placement and the light propagation path through the tissue.

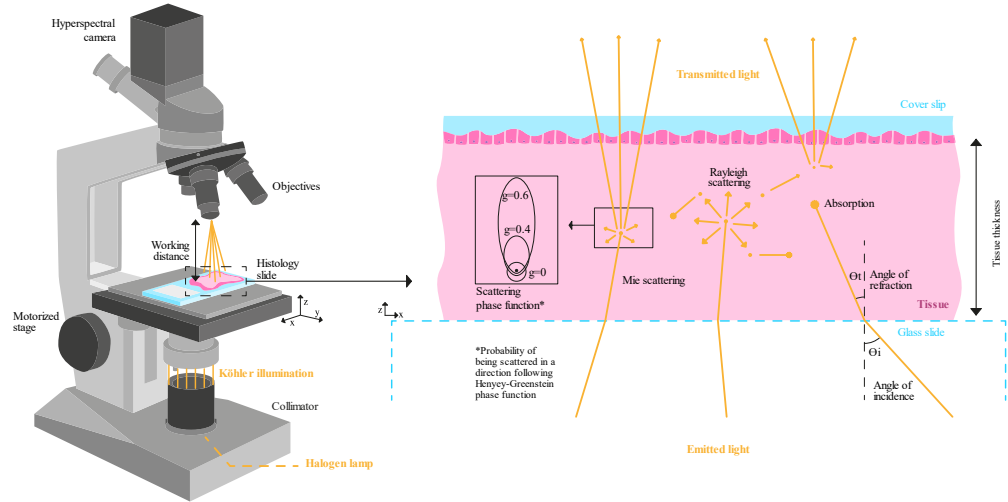

Fig. 2. Schematic of microscopic HS system capturing a histology slide in transmittance mode (left) and light-tissue interactions within the tissue being analyzed (right).

To eliminate the impact of the acquisition system on the HS images, a flat-field correction process is commonly applied to the raw data using both white and dark reference images [8]. In transmittance mode, a region of the transparent glass slide without specimen is used to capture the white reference spectrum ($White\ Ref$), while the dark reference spectrum ($Dark\ Ref$) is acquired by turning off the light source. The transmittance pixel ($Transmittance\ Pixel$) is then calculated using Eq. (1), with the raw pixel ($Raw\ Pixel$) as the initial data captured by the sensor. Through proper calibration, the transmission effects of the glass slide are compensated for. Consequently, in these simulations, the glass is not considered. Furthermore, the dark reference spectrum is not simulated, as sensor noise is not being simulated.

$$Transmittance\ Pixel = \frac{Raw\ Pixel - Dark\ Ref}{White\ Ref - Dark\ Ref} \quad (1)$$

*3.2 Tissue Optical Properties*

The next step in designing the MC framework is to define the optical properties of the tissues (more information in the Supplementary Material):

- The absorption coefficient ($\mu_a$) was computed, following Jaques [30], as a weighted sum of the main chromophores (oxygenated hemoglobin ($\mu_a.oxy$), deoxygenated hemoglobin ($\mu_a.deoxy$), water ($\mu_a.water$), and fat ($\mu_a.fat$)) according to Eq. (2):

$$\mu_a = BS\mu_{a.oxy} + B(1-S)\mu_{a.deoxy} + W\mu_{a.water} + F\mu_{a.fat} \quad (2)$$

Here, B is the blood volume fraction, S is the hemoglobin oxygen saturation, W is the water content, and F is the fat content. Minor absorbers such as melanin, bilirubin and β-carotene were excluded due to their minimal concentration in the tissues under study [30].

- The <u>reduced scattering coefficient ($\mu'_s$)</u> was modeled as a combination of Rayleigh and Mie scattering effects, with wavelength dependence relative to a reference wavelength $\lambda_{Ref} = 500\ nm$. The formulation is given in Eq. (3):

$$\mu'_s = a\left(f_{Ray}\left(\frac{\lambda}{\lambda_{Ref}}\right)^{-4} + (1-f_{Ray})\left(\frac{\lambda}{\lambda_{Ref}}\right)^{-b_{Mie}}\right) \quad (3)$$

where $a$ is a scaling factor, v is the Rayleigh contribution weight, and $b_{Mie}$ is the Mie scattering exponent.

- The <u>scattering phase function</u> describes the angular distribution of scattered photons. While the standard Henyey-Greenstein (HG) function, defined in Eq. (4), is commonly used:

$$p_{HG}(g,\theta) = \frac{1}{4\pi}\frac{1-g^2}{(1+g^2+2g\cos\theta)^{3/2}} \quad (4)$$

we employed the more accurate Two-Term HG model, Eq. (5), to better capture forward and backward scattering in thin tissue samples:

$$p(\alpha, g_f, g_b, \theta) = \alpha \cdot p_{HG}(g_f, \theta) + (1-\alpha) \cdot p_{HG}(g_b, \theta) \quad (5)$$

Here, $\alpha$ is the weight of forward scattering, and $p_{HG}(g_f, \theta)$ and $p_{HG}(g_b, \theta)$ are the anisotropy parameters for forward and backward scattering, respectively.

- The <u>refractive index (n)</u> was fixed at 1.35, consistent with values reported for similar biological tissues, and defined in (6) and (7):

$$n = \frac{c}{v} \quad (6)$$

$$n = \frac{\sin\theta_i}{\sin\theta_r} \quad (7)$$

where c is the speed of light in vacuum, v in the medium, and $\sin\theta_i$ and $\sin\theta_r$ are the angles of incidence and refraction, respectively.

### 3.2.1 Optical Properties of Tissue under Evaluation

A comprehensive review of the state-of-the-art literature pertaining to the eight tissues under investigation (breast, lung, colorectal, and liver, in both pathologies, lesioned and healthy) was conducted to extract their optical properties. Absorption and scattering parameters for different tissues were extracted from several literature sources. According to Eq. (2), the key parameters for calculating absorption are the percentages of blood (B), fat (F), and water (W), as well as blood oxygen saturation (S). For scattering calculations, the required parameters include the reference wavelength (λRef), Rayleigh scattering fraction (f_Ray), and Mie scattering parameters (b Mie and α). For breast tissue, data from L. L. de Boer et al. [31] were used. Fat content was estimated as 90% of the tissue volume excluding blood, with the remaining 10% assigned to water. Lung tissue parameters were obtained from J. W. Spliethoff et al. [32], who directly provided blood and water percentages; fat content was then calculated as the residual fraction. All values for colorectal tissue were sourced from M. S. Nogueira et al. [33]. For liver tissue, parameters came from N. Reistad et al. [34], reporting fat as 10.1% of the total volume excluding blood, with water making up the remaining 89.9%. Scattering phase function parameters (forward scattering (gf), backward scattering (gb), and anisotropy (α)) were also taken from multiple studies: N. Ghosh et al. [35] for breast, R. Marchesini et al. [36] for lung, A. N. Bashkatov et al. [37] for colorectal, and both P. Saccomandi et al. and R. Marchesini et al. [38,39] for liver.

Mean values of absorption, scattering, and phase function optical properties are presented in Table 1, with their standard deviations available in Table S1 of the Supplementary Material. Fig. 3 illustrates these mean values along with their corresponding standard deviations for the tissues under evaluation. It is important to note that, similar to regular tissue, simulated tissue on microscopy slides includes both fat and water, but these values might be different after histopathologic processing or frozen sectioning, where some water is lost. While these values are crucial in real-life scenarios for thin tissue slicing, they are included here to simulate pure, freshly tissue excised, and not processed tissue.

Table 1. Optical Properties parameters for normal and lesioned samples of breast, lung colorectal, and liver tissues.

| | | Absorption | | | | | Scattering | | | R | Scattering Phase Function | | | R |
|---|---|---|---|---|---|---|---|---|---|---|---|---|---|---|
| | | B (%) | S (%) | F (%) | W (%) | λRef (nm) | b Mie | f_Ray | a | | gf | gb | α | |
| Breast | N | 1,75 | 16* | 88,43 | 9,83 | 800 | 0,69* | 5 | 14 | [31] | 0,87 | -0,09 | 0,82 | [35] |
| | L | 3,5 | 41* | 86,85 | 9,65 | 800 | 0,69* | 15 | 25 | | 0,92 | -0,06 | 0,84 | |
| Lung | N | 7,5 | 90 | 35,5 | 57 | 800 | 0,98* | 0* | 38 | [32] | 0,84 | -0,5 | 0,92 | [36] |
| | L | 2,5 | 69 | 3,5 | 94 | 800 | 0,98* | 0* | 21 | | 0,84 | -0,5 | 0,92 | |
| Colorectal | N | 5 | 64 | 7,7 | 78,5 | 500 | 0,2 | 56 | 14,3 | [33] | 0,89 | 0 | 1 | [37] |
| | L | 6,6 | 66 | 1,4 | 84,9 | 500 | 0,5 | 40 | 15,9 | | 0,89 | 0 | 1 | |
| Liver | N | 8,5 | 50 | 9,24 | 82,26 | 800 | 0,35 | 29 | 13,7 | [34] | 0,88 | -0,46 | 0,93 | [38,39] |
| | L | 7,2 | 50 | 9,37 | 83,43 | 800 | 0,05 | 10,7 | 18,2 | | 0,88 | -0,46 | 0,93 | |

N: Normal tissue, L: Lesioned tissue, B: Blood, S: Saturation, F: Fat, W: Water, R: Reference. *Extracted from [30].

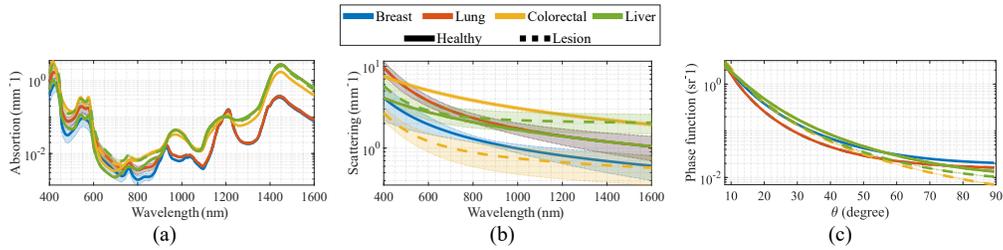

Fig. 3. Optical properties of tissues (breast, lung, colorectal, and liver) in both normal and lesioned states: (a) absorption coefficient, (b) scattering coefficient, and (c) scattering phase function.

### 3.3 MC Simulation Framework

To investigate light–tissue interactions under controlled and reproducible conditions, a custom MC simulation framework was developed. This framework was developed to simulate the HS imaging system described in Section 3.1, producing synthetic radiance data based on well-defined tissue conditions. This approach enables systematic analysis of how key parameters, such as tissue thickness, affect the resulting spectral signatures. An overview of the complete simulation pipeline is presented in Fig. 4.

### 3.3.1. Simulated Volume

The MC simulation was set up to emulate a homogeneous single-layer sample with a spatial extent of 100×800 μm. To evaluate the influence of tissue thickness on the spectral response, he height of the simulated volume was varied across several depths: 5, 20, 50, 100, 200, 500, 1000, and 2000 μm. These values were selected to span from typical histological sections (5 μm) to depths where light transmittance becomes negligible due to increased tissue attenuation (2000 μm). However, simulating thick samples at high spatial resolution can be computationally intensive. To manage this, the voxel size and the number of voxels were adjusted according to the tested thickness:

- For 5 and 20 μm thicknesses, a voxel size of 1 μm was used, yielding volumes of [5, 20] × 100 × 800 voxels (depth, lines, pixels).

- For 50, 100, and 200 μm thicknesses, a voxel size of 10 μm was used, resulting in volumes of [5, 10, 20] × 10 × 80 voxels.

- For thicker samples (500, 1000, and 2000 μm), a voxel size of 100 μm was used, leading to volumes of [5, 10, 20] × 1 × 8 voxels.

This adaptive resolution strategy ensured that the number of simulated voxels remained within a manageable range, while still allowing for meaningful comparisons across different thicknesses.

### 3.3.2. Simulated Tissue

For each tissue type (breast, lung, colorectal, and liver) and its corresponding pathological states (normal, lesioned), the mean and standard deviation of the optical properties were collected from the literature (see Table I from Supplementary Material). To introduce variation within each tissue type, and assuming a normal distribution, ten random values were selected for each tissue–pathology combination. In total, 80 biological tissues were simulated (10 per tissue–pathology combination).

### 3.3.3. Simulated Spectra

Replicating the HS microscopy setup described in Section 3.1, each MC simulation covered the 400–1000 nm spectral range. A total of 105 spectral bands were simulated, corresponding to a distance between bands of 5.7 nm. While different number of bands with different spectral resolutions could have been selected to better represent a specific MS or HS application, 105 bands were chosen to provide a balanced and representative sampling of the spectral range defined.

### 3.3.4. Flat-field Correction

Flat-field correction is a critical preprocessing step in HS imaging, used to compensate for non-uniformities in illumination, sensor response, and optical path variations [40,41]. To simulate the flat-field reference ($I_0$), the MC simulation was run without any tissue in the optical path, emulating the acquisition of an empty region on a microscope slide. Although photons were propagated through air (assumed to have negligible absorption and scattering) not all were expected to reach the detector. Instead, the simulation estimated the number of photons passing through the sensor slit at the detector's working distance (18 mm).

Next, the optical properties were replaced with those of biological tissue (e.g., breast tissue), and the corresponding radiance spectra ($I$) were generated across the wavelength range. Flat-field correction was then applied according to Eq. (8) by dividing, at each wavelength, the number of photons transmitted through the tissue by the ones detected in the absence of a sample. This process yielded normalized tissue transmittance values.

It is important to note that, since all simulations were conducted on spatially homogeneous volumes and no spatial variation was modeled, spatio-spectral calibration (typically required in real-world HS imaging systems) was not necessary in this study.

$$T = \frac{I}{I_0} \quad (8)$$

### 3.3.5. Software and Hardware Specifications

A total of 67,200 simulations were performed (8 tissue thickness × 80 biological tissues × 105 spectral bands), with each iteration simulating $10^8$ photons. Due to the high computational demand, MC simulations were executed using Monte Carlo eXtreme (MCX) version 2023.11, developed by Qianqian Fang [42–45], which supports GPU acceleration. Simulations were run through MATLAB R2023a on a workstation equipped with an Intel Xeon Silver 4216 (16-core) CPU, 128 GB of RAM, and three NVIDIA Tesla T4 (TU104GL) GPUs.

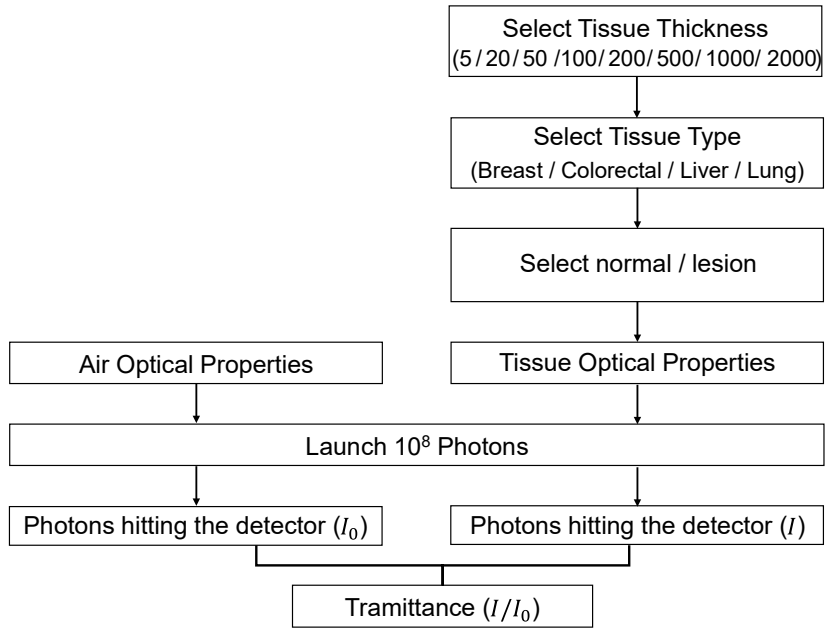

*Fig.* 4. Framework followed to perform the different MC simulations obtaining the spectral signatures of various organ tissues, at two tissue types at several tissue thicknesses.

### *3.4 Data Augmentation*

To establish a realistic database, in alignment with real-world samples (e.g. including the sensor signal-to-noise ratio, wavelength miscalibration, improper light source thermal management, or intensity fluctuation due to changes in the light source), data augmentation was performed over the simulated spectral signature database. The procedure outlined in [46] was followed, where each stage of the process builds upon the previous one.

First, two wavelength miscalibration processes were simulated. The spectra's wavelength range experienced reduction or extension by a random value ranging from -1 to 1 nm, with the spacing between bands adjusted in concordance. For instance, if the original vector spanned from 400 to 1000 nm with 100 bands and now is extended from 400 to 1001 nm, the spacing between bands is augmented from 6 to 6.01 nm. The second wavelength miscalibration technique was to shift the wavelength vector within the range ± 4.8 nm. This would represent measurements taken on different days, which assessed the shift in light spectra. Finally, the spectra were interpolated to the original wavelength vector to have all the data in the same wavelength vector.

Following these adjustments, two spectra intensity errors were introduced simulating fluctuations in light source intensity [46]. The first error involved wavelength-dependent intensity fluctuations, where the spectrum's intensity was randomly tilted following a linear shape, by up to 5 % of the spectral signature mean. For instance, a spectral signature with a mean 0.8 would potentially be scaling the left tail up to 0.84 and the right tail down to 0.76. The second intensity error involves a constant intensity shift of up to 5 % of the spectra mean.

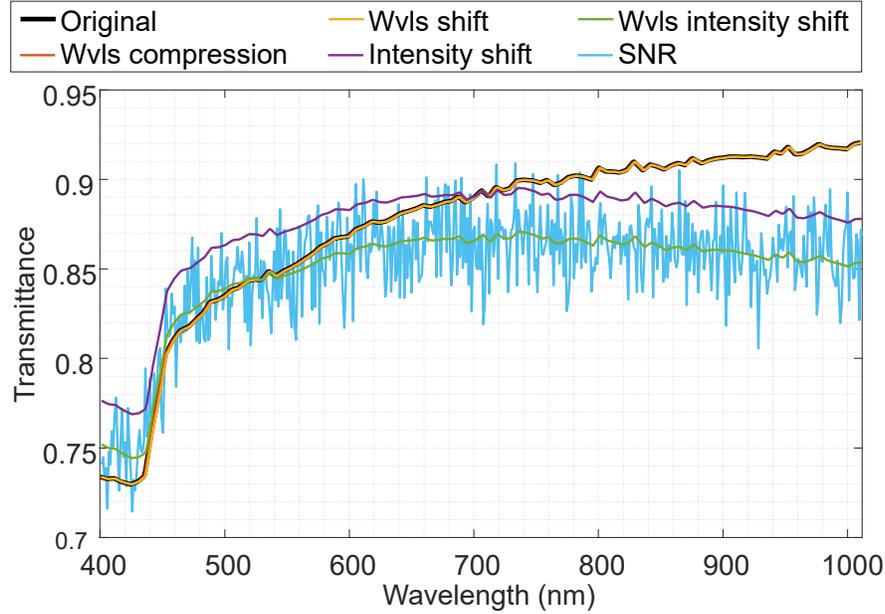

Fig. 5. Data augmentation applying different strategies to the original spectral signature database. The original transmittance spectrum is altered through wavelength (wvls) compression, wavelength shift, intensity shift, wavelength intensity shift, and finally white Gaussian noise is added.

In the final step, noise was introduced to the spectra to replicate the sensor's intrinsic noise floor. First, the dataset was expanded by interpolating five additional points between each simulated band pair, increasing the spectral resolution from 105 to 521 bands. This expansion provided a larger number of data points on which the noise could be applied. Additive white Gaussian noise was then applied to simulate a signal-to-noise ratio (SNR) of 35 dB, simulating the noise of a real HS microscopy system. Fig. 5 visually represents the original spectral signature, which refers to the one obtained using the MC simulation framework, and each stage of the data augmentation process.

*3.5 Spectral Evaluation Metrics*

Several spectral evaluation metrics were employed to assess the differences between the two spectral signatures. Let $t$ be the test spectra, $r$ the reference spectra and $N$ be the number of bands, the following metrics are as follow:

3.5.1 Euclidean Distance

The Euclidian distance ($d$) quantifies the distance between two vectors in an N-dimensional space. It is determined by Eq. (9), which computes the quadratic mean of variances between test and reference values. The Euclidean distance is bounded between 0 and 1 (for $t$ and $r$ in the [0, 1] range), with a value close to zero indicating a high similarity between the spectral signatures.

$$d = \sqrt{\sum_{i=1}^{N}(t_i - r_i)^2} \qquad (9)$$

### 3.5.2 Spectral Angle Mapper

The Spectral Angle Mapper (SAM) assesses spectral similarity by computing the angle between spectra, treated as vectors in an N-dimensional space [47]. It is calculated using (10). This metric represents angles, so it ranges from 0 to $\pi/2$, when both, $t$ and $r$ are non-negative. A smaller SAM value indicates greater spectral similarity.

$$SAM = \arccos\left(\frac{\sum_{i=1}^{N} t_i r_i}{\sqrt{\sum_{i=1}^{N} t_i^2} \sqrt{\sum_{i=1}^{N} r_i^2}}\right) \qquad (10)$$

### 3.5.3 Normalized Spectral Similarity Score

The Normalized Spectral Similarity Score (NS$^3$) is a method for evaluating spectral similarity by considering both Euclidean and SAM distances between spectra. It offers a novel approach to spectral comparison by addressing potential misidentifications arising from ambiguous high-confidence scores in spectral amplitude and angle. Following (11), NS$^3$ normalizes spectral amplitudes and applies a custom function to match the spectral angle and amplitude difference scores, enhancing accuracy. Taking into account the ranges of $d$ and $SAM$, this metric ranges from 0 to $\sqrt{2}$, where a low NS$^3$ score indicates a strong correspondence between the test and reference signatures.

$$NS^3 = \sqrt{d^2 + \left(1 - cos(SAM)\right)^2} \qquad (11)$$

### 3.5.4 SID

Spectral Information Divergence (SID) is an information-theoretic spectral measure introduced to assess the dissimilarity between the spectral signatures of two pixels in HS images [48]. It is developed based on the concept of divergence, specifically measured in the probabilistic behavior between the spectral characteristics of the compared pixels (12). SID values lie in the range $[0, \infty)$, where a smaller SID indicates low divergence between the spectral signatures of the compared pixels.

$$SID = \sum_{i=1}^{N} t_i \, log\left(\frac{t_i}{r_i}\right) + \sum_{i=1}^{N} r_i \, log\left(\frac{r_i}{t_i}\right) \qquad (12)$$

### 3.5.5 SID-SAM

The SID-SAM mixed measure, calculated according to Eq. (13), leverages the strengths of both SID and SAM in spectral discriminability [49]. This implies that the spectral similarity and dissimilarity achieved through the mixed measure are significantly improved by multiplying the spectral capabilities of both measures. The choice of tangent or sine over cosine is made to compute the perpendicular distance between two vectors, rather than calculating the projection of one vector along the other. A SID-SAM score close to 0 signifies a robust match between the test signature and the reference signature and tends to infinity when the maximum dissimilarity is reached.

$$SID(TAN) = SID \times \tan(SAM) \qquad (13)$$

### *3.6 Quantitative Thickness Evaluation*

The previously presented evaluation metrics were calculated to measure the difference between inter-class (normal (N) vs lesion (L)) and intra-class (N vs N and L vs L) spectra. Fig. 6 (a) shows the Euclidean distance of inter-class (orange histograms) and intra-class (blue histograms) liver spectra. To obtain a metric that yields consistent results within the same class while emphasizing differences for spectra from different classes, the disparity between the inter-class and intra-class histograms for a given thickness serves as a qualitative indicator of the discrimination capability associated with that thickness.

However, a quantitative metric is needed to measure this discrimination. At each thickness, a threshold ($th$) can be defined to categorize the inter and intra-class distances as belonging to inter or intra-class evaluations. The accuracy metric can then be calculated over the true positive, true negative, false positive, and false negative values. Under a specified $th$, the accuracy of the classification is determined by (14), where each $bin_i$ represents the individual values within a histogram. Since spectra within a class are more similar to each other (intra-class distances) than to the spectra of the other class (inter-class distances), intra-class distances should be lower than the inter-class ones. Therefore, the distributions are ordered such that values below the $th$ are classified as intra-class, while values above it are classified as inter-class, providing most of the time accuracy values between 0.5 and 1. In scenarios where histograms overlap, the selection of any $th$ would yield a substantial number of misclassifications, indicating a constrained discriminative capacity in these thicknesses (e.g., for Fig. 6 (b) the threshold providing maximum accuracy (0.54) was found at an Euclidean distance of 0.05). Conversely, in thicker samples, where histograms manifest greater separation, establishing a $th$ becomes more straightforward (e.g., for Fig. 6 (c) the threshold providing maximum accuracy (0.83) was found at an Euclidean distance of 0.07).

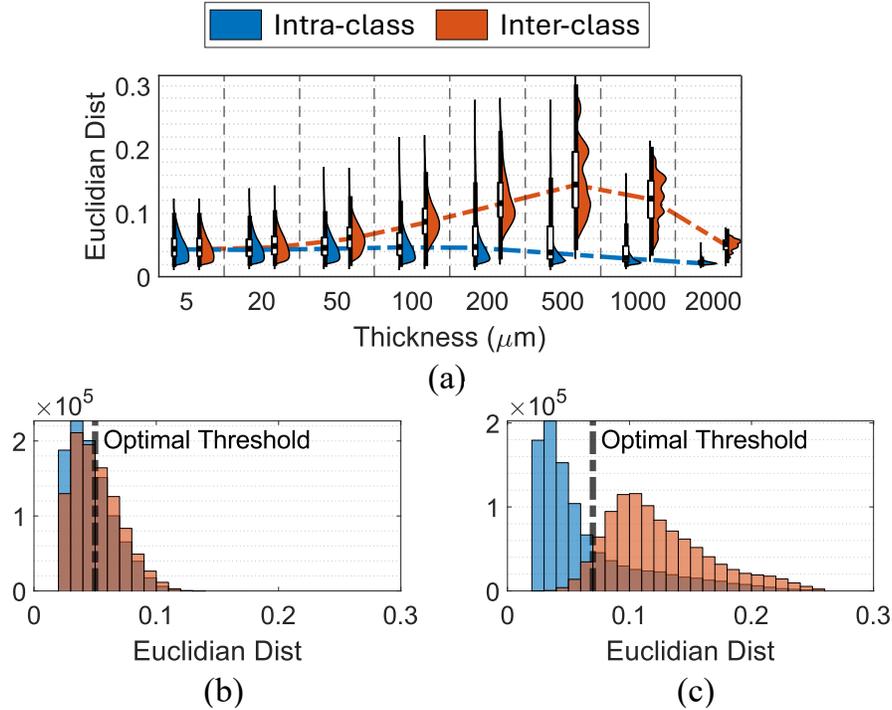

Fig. 6. (a) Euclidean distance scores between intra-class (blue) and inter-class (orange) spectra at different tissue thicknesses for liver tissue. Zoom over the histograms at (b) 20 μm and (c) 200 μm, red line showing threshold with maximum accuracy for each case.

The accuracy was calculated across all possible $th$s for a single set of histograms, covering the lowest to the highest obtained metric scores, with an increment of one bin at a time. The highest accuracy value obtained across all thresholds was chosen to be the discriminative power of that thickness. Subsequently, the methodology was repeated for all combinations of tissue type (breast, colorectal, liver, and lung), metric (Euclidean distance, NS3, SAM, SID, SIDSAM), and thickness (5, 20, 50, 100, 200, 500, 1000, and 2000 μm). This process allowed us to identify, at each tissue and metric, the thickness that yielded the best accuracy between lesion and normal tissue.

$$ACC = P(intra-class) \sum_{i=1}^{th} P(bin_i|intra-class) + \\ P(inter-class) \sum_{i=th}^{end} P(bin_i|inter-class) \qquad (14)$$

## 4. Results

In this section, the results obtained following the previous methodology are presented. The MC simulated dataset after data augmentation is shown in Fig. 7 for four of the studied thicknesses (all thicknesses shown in Fig. S1 in the Supplementary Material), where green and red spectra correspond to normal and lesioned tissue, respectively. As anticipated, tissue samples with a thickness of 5 µm exhibit flat transmittance spectra, with values approaching 1. This suggests that there are no distinct peaks associated with the absorption or scattering of endogenous chromophores within the tissue. These results confirm that samples thinner than the diffusion length (6 to 12 mm [50]) do not experience significant scattering or absorption. At this thickness, there is no notable difference in the spectral signatures of normal and lesioned tissues. However, as the sample thickness increases, the distinction between normal and lesioned tissue becomes more apparent. By 50 µm, differences in spectral signatures between tissue types start to emerge, indicating that conventional sample thicknesses of 2 to 10 µm may not provide sufficient contrast for pathology discrimination with HS microscopy. At 500 µm, the contrast between tissue types is most pronounced, although the mean transmittance drops below 0.5, a value that could be even lower in practical applications due to instrumental attenuation. Beyond this thickness, light penetration diminishes significantly, reducing the signal that reaches the sensor. The maximum simulated thickness of 2000 µm is well beyond the cutting capacity of standard microtomes, which are typically limited to around 60-70 µm. At these thick samples (>2000 µm), the spectral signatures appear nearly flat due to minimal light transmission through the sample, resulting in transmittance values approaching zero across all tissues and pathologies. This results in a lack of spectral resolution at this thickness.

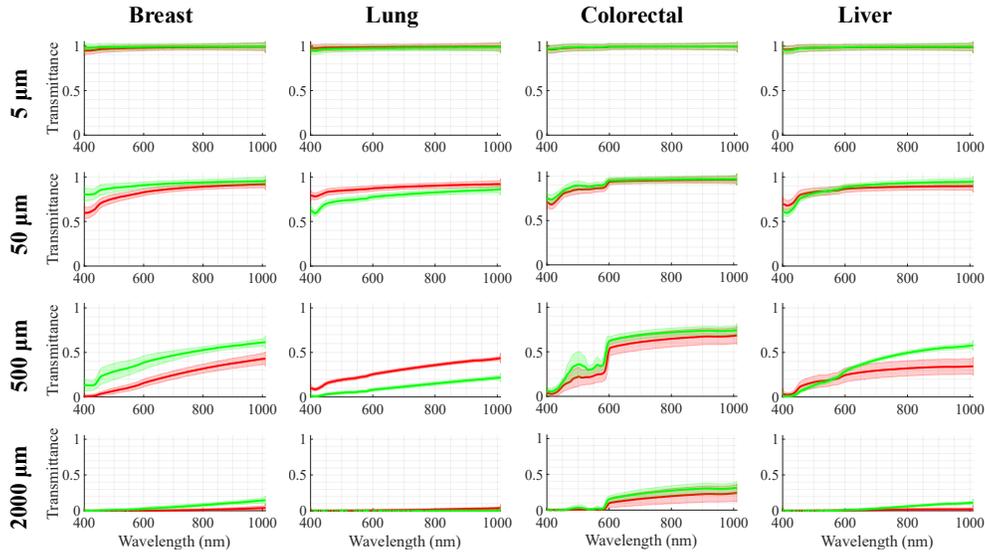

Fig. 7. Spectral signatures for normal (green) and lesion (red) tissues for each type of tissue and four slide thickness

After simulating the entire dataset, the evaluation of the data was performed using the metrics described in the 3.5 Spectral Evaluation Metrics section. Inter- and intra-class distances were calculated using Euclidean distance, SAM, NS3, SID, and SID-SAM (see Fig. S2 from

Supplementary Material). Subsequently, in accordance with the methodology described in the 3.6 Quantitative Thickness Evaluation, the aim was to assess the discrimination power of each tissue thickness. This evaluation focused on identifying which tissue can be accurately distinguished between normal and lesioned tissues (inter-class spectra) while also accurately classifying them as equivalent when they belong to the same class (intra-class spectra). Fig. 8 illustrates the accuracy results for each tissue type and metric.

Different metrics exhibit varying levels of discrimination due to their inherent characteristics (e.g., SAM measures the relative difference between spectral signatures, while Euclidean distance is an absolute metric). However, all metrics consistently show that a thickness of 5 µm offers no discrimination between pathologies on unstained samples (accuracy around 0.5). The optimal discrimination thickness was determined by first identifying the thickness that yielded the highest accuracy for each evaluation metric individually. A majority voting approach was then applied to select the thickness at which the greatest number of metrics reached their maximum performance. This approach is metric-agnostic, as not all metrics may be equally relevant for a given application, allowing flexibility in selecting the most appropriate metrics based on the specific requirements of the intended use case. Based on this method, the optimal discrimination thicknesses were found to be 1000 µm for breast tissue, 200 and 500 µm for lung and colorectal tissues, and 500 µm for liver tissue.

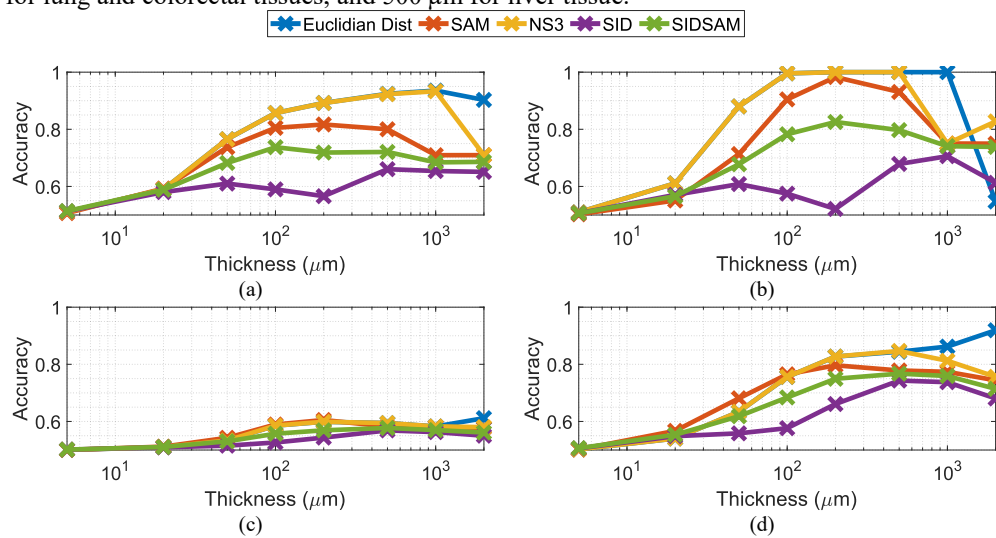

Fig. 8. Spectral difference between tumor and normal tissues fusing (a) breast, (b) lung, (c) colorectal and (d) liver. Thickness is represented on the x-axis on a logarithmic scale.

## 5. Discussion

In recent years, there has been a growing adoption of MS/HS microscopic systems. However, it is essential to exercise caution in their application. In traditional histology, samples typically range from 2 to 10 µm in thickness, which is thinner than the diffusion length in tissue (6 to 12 mm [50]). Given the absence of absorption and scattering in these thin samples, dyeing becomes necessary to produce image contrast. However, employing sample dyeing in MS/HS microscopic imaging is suboptimal as the final goal is to capture information regarding the interaction of light with intrinsic biomarkers present in tissue (endogenous chromophores). An in-depth analysis of the current state-of-the-art was conducted to find out the tissue thickness utilized in previous MS/HS microscopic studies. From this review, it was found that 60% of the documents did not report the tissue thickness employed in the experiments, and from the other 40 %, all works but one analyzed samples of 2 to 10 µm thicknesses, following traditional histology procedures. The objective of this work was to assess how the choice of sample thickness affects spectral contrast on MS/HS microscopy data and to suggest an approach for understanding the impact of thickness on the resulting spectral signatures.

A MC Light Transport Simulator framework was developed to investigate tissue-light interactions at various thicknesses (5, 20, 50, 100, 200, 500, 1000, and 2000 μm) in both normal and lesioned tissues, including breast, lung, colorectal, and liver. This process involved a thorough analysis of a microscopic HS system, with particular attention to its instrumentation and optical components. Furthermore, the optical properties of the tissues, such as absorption, scattering, scattering phase function, and refractive index, were sourced from the literature to create a virtual model of the tissues for the simulations. After the development of the MC framework, ten values were selected for each tissue, pathology, and thickness, making a total of 640 simulated spectral signatures in the range of 400 to 1000 nm. Once the simulations were completed, data augmentation techniques were applied to the original dataset to better reflect real-world HS microscopic conditions. These techniques simulated practical factors such as sensor SNR, wavelength calibration errors, inadequate thermal management of the light source, and intensity fluctuations due to variations in the light source, thereby enhancing the simulation data with more realistic scenarios.

MC simulated spectra from the virtual specimen illustrated that conventional histology thicknesses (~5 μm) would produce uniform light responses, suggesting that contrast in standard histology images primarily originates from dye absorbance (exogenous chromophores). Conversely, excessively thick samples (1000 and 2000 μm) would block light transmission, resulting in limited photon capture by the HS camera. Qualitatively, samples sliced at 500 μm demonstrated enhanced discrimination between normal and lesioned tissue, although they exhibit a relatively low maximum intensity (50 % of the maximum achievable). To quantitatively evaluate the discrimination factor of each thickness, at which normal and lesioned tissue become distinguishable, several spectral evaluation metrics were employed (Euclidean Distance, SAM, $NS^3$, SID, and SID-SAM). For each thickness, these metrics evaluated all spectral signatures between them, categorizing the results into two groups: inter-class (N vs L) and intra-class (N vs N | L vs L) comparisons. The goal was to identify the thickness where a given metric best discerned between spectra from different classes, characterized by small intra-class distances and large inter-class distances. For each thickness, histograms of inter-class and intra-class distances were created, and a range of thresholds were tested to find the one providing the most accurate classification of classes. The accuracy of the classification at the optimal threshold for each thickness indicated the discrimination power of the spectral evaluation metric at that thickness. As expected from the qualitative analysis, results showed that a thickness of 5 μm did not provide adequate differentiation between spectra from the same or different classes. For other thicknesses, each tissue type exhibited a distinct maximum discrimination thickness: 1000 μm for breast tissue, 200 and 500 μm for lung and colorectal tissues, and 500 μm for liver tissue.

However, caution is required when interpreting the absolute values presented in this paper since several limitations must be considered in this study. The simulated tissue closely approximates in vivo conditions (based on the available values in the current literature), although histological analysis is performed on ex vivo samples. The biopsy procedure entails cutting and slicing, which results in blood loss from the sample, leaving an arbitrary residual volume. Furthermore, the remaining hemoglobin interacts with oxygen in the air, forming oxyhemoglobin. As a result, the observed saturation levels may approach 100%, making them irrelevant to the original tissue saturation. Additionally, the preparation of formalin-fixed stained slices may lead to the loss of water and fat. Other endogenous chromophores may also be missing, or the homogenized structure of the sample may not accurately resemble the original tissue. These factors should be considered in future simulations tailored to specific applications. Moreover, these simulations do not account for the spectral signatures associated with sample fixation methods involving chemicals such as formalin and paraffin. Variations in instrumentation (such as power loss in the optical system and the effects of the histology glass holding the samples) have also not been simulated. It is also important to note that the simulations were performed on a spatially homogeneous sample, without accounting for structural heterogeneity and its effects on the resulting MS/HS image (e.g., optical vignetting, image aberrations, focus variation, etc.). Thus, while the virtual specimen enables the simulation of light transmission through various tissues and thicknesses, the absolute spectral signature values obtained are not expected to perfectly reflect real-life scenarios. Nonetheless, the results of the simulation show that for unstained hyperspectral microscopy, the tissue thickness is a factor that should be considered during sample acquisitions.

## 6. Conclusions

In conclusion, this study underscores the critical importance of carefully considering sample preparation protocols in MS/HS microscopic applications, as conventional histological methods may not be able to provide the enhanced information derived from light-tissue interactions in thicker samples. The simulated spectral signatures presented here provide valuable insights into how light-tissue interactions vary with tissue thickness, serving as a useful reference point for future studies. However, further research is needed to determine the optimal sample thickness for each tissue type to ensure more accurate and reliable results in real-world applications. The methodology developed in this work can be adapted and extended to other systems and tissue types, allowing researchers in the field to identify the optimal thickness for their specific applications. Different MS/HS sensors with varying bandwidths and numbers of spectral bands may also be simulated. The discriminative power of each thickness depends on the tissue composition (which will be heterogeneous and more complex than the single-layer model proposed in this paper) and on the quality of the spectral signatures (which will depend on the instrumentation). Nonetheless, a correlation between the simulations and the experimental results is expected when the spectral resolution of the instrumentation used in the experiments aligns with the simulated one. However, this correlation must be validated empirically.

Thus, while simulated spectra offer insights into which tissue thicknesses provides enhanced spectral contrast, future studies should empirically validate these findings. The framework developed in this work can be used to perform a finer search for the optimal thickness of a specific tissue, which can then be validated using HS microscopy data acquired from real tissue sections. Validation would involve classification tasks, such as distinguishing between different tissue types (e.g., liver vs. lung) or pathological states (e.g., normal vs. tumor), to assess whether the selected thickness enhances classification performance. Afterwards, it would be crucial to explore how variations in sample thickness can be effectively achieved and incorporated into the clinical workflow. While the aim is not to drastically alter existing clinical processes, if unstained thicker samples lead to improved spectral data in MS/HS microscopy, their integration into clinical procedures could enhance diagnostic accuracy beyond the current state of the art.

**Funding.** Consejería de Educación, Universidades, Cultura y Deportes, Gobierno de Canarias (Research Stay, EST2023010017); Agencia Canaria de Investigación, Innovación y Sociedad de la Información (TESIS2021010084); European Health and Digital Executive Agency (101137416); Ministerio de Ciencia, Innovación y Universidades (PID2023-148285OB-C43).

**Acknowledgment.** We gratefully acknowledge Óscar Quintana for his work in developing Fig. 2 presented in this study. His expertise in data visualization was instrumental in enhancing the clarity and impact of this work. The authors also thank Guillermo V. Socorro-Marrero for his support in the statistical analysis of this work.

**Disclosures.** The authors declare no conflicts of interest.

**Data availability.** Data underlying the results presented in this paper are not publicly available at this time but may be obtained from the authors upon reasonable request